\documentclass{article}
\usepackage{graphicx}
\usepackage{dcolumn}
\usepackage{rotating}
\usepackage{supertabular}
\usepackage{longtable}
\usepackage{bm}
\usepackage{verbatim}
\usepackage{textcomp}
\usepackage{amsfonts}
\usepackage{amssymb}
\usepackage{amsmath}
\usepackage{listings}
\usepackage{subfigure}

\title{A Minimum Variance Method for \\ Problems in Radio Antenna
Placement\thanks{A short version of this work appears in the
proceedings of The Low Frequency Radio Universe, ASP Conference
Series, Vol LFRU, 2009, Eds: D.J. Saikia, Dave Green, Y.Gupta and
Tiziana Venturi. Email: mvpandurangarao.m@tcs.com} }
\author{M. V. Panduranga Rao, Amrit Lal Ahuja, Srinivasan Iyengar, Sachin Lodha
\\ \small{Tata Research Development and Design Centre, Pune, India}\and
Kavita Iyer, Ranu Khade,\\ \small{College of Engineering, Pune,
India} \and
 Dinesh Mehta, Balasz Nagy\\ \small{Colorado School of
Mines, Golden, USA} \\~\\}

\begin{document}

\date{\today}
\maketitle

\begin{abstract}
Aperture synthesis radio telescopes generate images of celestial
bodies from data obtained from several radio antennas. Placement of
these antennas has always been a source of interesting problems.
Often, several potentially contradictory objectives like good image
quality and low infra-structural cost have to be satisfied
simultaneously.

In this paper, we propose a general Minimum Variance Method that
focuses on obtaining good images in the presence of limiting
situations. We show its versatility and goodness in three different
situations: (a) Placing the antennas on the ground to get a target
Gaussian UV distribution (b) Staggering the construction of a
telescope in the event of staggered budgets and (c) Whenever
available, using the mobility of antennas to obtain a high degree of
fault tolerance.

\end{abstract}

\section{Introduction}      \label{sec:intro}
The technique of interferometric aperture synthesis has brought
about a revolution in the field of radio astronomy in the past four
decades\cite{Ryle}. The technique works by synthesizing images from
signals obtained from antennas spread over a large distance. For a
quick introduction to basic interferometric radio astronomy, please
see the Appendix A.

Most radio telescopes today use the concept of interferometric
aperture synthesis. The Very Large Array (VLA) in the USA with 27
antennas and the Giant Meter wave Radio Telescope (GMRT) in India
with 30 antennas are good examples of such telescopes. A
multi-purpose new generation radio telescope called the {\it Square
Kilometer Array} (SKA) has been proposed (see Section
\ref{sec:prev_wrk} and the website http://www.skatelescope.org), and
is currently in the design phase.

An important problem in the field of interferometric radio-astronomy
is to find the antenna placement on the ground that generates a
required UV-distribution.\footnote{For more on the terminology and
concepts, please see Appendix \ref{app:primer}.} This problem is a
computationally difficult one~\cite{MOORTA}. In addition to antenna
placement on purely scientific merit, other dimensions are added
when we consider logistical and financial issues. For example, given
a particular placement, what is the optimum cable layout? What
trade-off can be struck between scientific merit and wire length and
other infrastructure costs?

In this paper we look at three important problems that arise chiefly
from the standpoint of astronomic merit:
\begin{itemize}
\item The first is a general problem. It turns out that a Gaussian distribution of UV points in the radial
direction and a uniform distribution in the azimuthal direction is
desirable for Gaussian beams~\cite{Boone2002}. It has been observed
empirically that a uniform random placement of antennas on the XY
plane yields a radially tapering UV distribution~\cite{Treloar}. A
simple calculation shows that this is as expected. However, it may
not be sufficiently close, particularly when some antenna positions
are not feasible (say, because of geopolitical constraints). How
does one rectify this?
\item The second is motivated primarily by financial
considerations. Suppose the funding for the project is spread over a
period of time as is likely to be in the case of the SKA. It is also
reasonable to assume that the construction of the entire array will
take time, because of logistical reasons. In such a scenario, one
would still like to perform experiments and get good quality images
in the meantime. Can we come up with a placement schedule that
enables such a graceful upgrading?

This question has been raised in the past in this
context~\cite{MOORTA} and others~\cite{Takeuchi}. A more recent
version of this question concerns expanding existing
arrays~\cite{Karast}. Their solution involves evaluating every
antenna configuration. This is expensive in the scenario where the
number of antennas is large, and there is limited time for
rearrangement of antennas between successive experiments.
\item The third concerns mobile antennas. While there are a limited number $M$ of antennas available, there
are several more pads that can host an antenna each. Ideally, an
experiment would require antennas on at least some of those pads.
Consider the case when only some $N$ of those antennas are available
(due to maintenance reasons or involvement with other experiments,
the rest are unavailable). On which $N$ pads out of the required
pads should we place the antennas?
\end{itemize}

We propose a Minimum Variance Method (MVM) that tackles the above
problems. It is interesting to note that the same framework is
useful for these seemingly different problems.

Informally, this technique involves choosing $N$ out of $M$ (where
$M\geq N$) possible locations for placing the antennas. To achieve
this, we start by placing an antenna on each of the M locations.
Then, we iteratively remove $M-N$ antennas, one at a time, such that
we stay ``close" to the target UV distribution. Our solution takes
$O((M-N)\cdot M^3)$ time, which is an improvement over brute force
solutions. A brute force solution would involve comparing
distributions that result from all $\binom{M}{M-N}$ choices for
antenna removal.

We conduct our experiments on a random placement scheme, on the
placement data of 120 antennas/stations of the proposed Australian
SKA (see next section), and on the placement data of the 24 antennas
of the Sub-millimeter Array at Mauna Kea, Hawaii (see Section
\ref{sec:app}).

Some relevant literature is discussed in the next section. In
Section \ref{sec:mvm}, we present the Minimum Variance Method.
Section \ref{sec:app} shows how the technique applies to the three
problems stated above, along with the experimental results. Section
\ref{sec:concl} concludes the paper.

\section{Previous Work}     \label{sec:prev_wrk}
This work has been done keeping in mind the proposed Square
Kilometre Array (SKA) project.  The SKA is an ambitious
multi-purpose new generation radio telescope with a total collecting
area of approximately one square kilometer, designed to work over a
wide range of radio frequencies -- $70 ~\rm MHz$ to $25 ~\rm GHz$.
The telescope is expected to play a major role in answering key
questions in modern astrophysics and cosmology. This high resolution
array will be $50$ times more sensitive, and will be able to cover
the sky $10000$ times faster than any imaging radio telescope array
previously built. It is hoped that besides exploratory astronomy,
the SKA will provide insights into many interesting questions about
the birth and evolution of galaxies, origins of magnetism,
possibility of existence of life on some other planet, verification
of general relativity etc. For more details see~\cite{Carilli}.

A lot of work has been done in the past on the problem of antenna
placement. They range from constructing the array ab-initio  to
incrementing an existing one. We cite a few important example papers
here; the reader is encouraged to read the citations therein.

In 1989, Treloar compared various array configurations in terms of
the amount of sampling of different regions in the UV-plane and the
absence of holes therein, for fixed declinations~\cite{Treloar}.
Among the configurations studied was the spiral, with higher
concentration towards the centre with distribution of antennas
varying with the radius. Cohanim et al posed the array design
problem as a multi-objective optimization problem of maximizing
image performance and minimizing cable-length using genetic
algorithms and simulated annealing~\cite{MOORTA}. Motivated by the
work of Takeuchi et al~\cite{Takeuchi}, they mention the problem of
phased deployment of the antennas.

Lonsdale and Cappallo make the case for a large number of antennas,
of the order of a thousand~\cite{Locap}. Among other factors
including high fidelity and good UV-coverage, they argue that a
large number of stations (a collection of antenna imaging a specific
region of the sky) relieves the dependence on the earth's rotation
for a comprehensive sampling of the UV-plane. They also present
log-spiral and hybrid (log-spiral for inner regions and
pseudo-random star for the outer regions) antenna layouts, which
have the advantages of good UV-coverage towards the centre, a wide
range of baseline lengths and shorter communication cables in
general. Problems of insufficient coverage and long cable lengths
appear in the outer regions. Several variants of a technique that
seeks to maximize the distance between UV points have also been
explored in the past (\cite{Boone2001} and~\cite{Cornwell88}).

Karastergiou et al~\cite{Karast} adapted the approach of
Boone~\cite{Boone2001} and Cornwell~\cite{Cornwell88} to the case
when there are more potential sites (called pads) than antennas, and
one has to choose a configuration for an experiment by shifting the
antennas among the pads. The work stemmed from an observation of
Boone~\cite{Boone2002} that if the density of the UV-plane is
sparse, it is advisable to spread the Gaussian along a radial
direction a bit so that all regions of the UV-plane have at least a
sample point. The technique they used was inspired by the physical
phenomenon of charges spreading on a closed 2-D surface in order to
minimize the energy. Their idea is to affect all the discrete shifts
and see which configuration results in the least energy, thus
satisfying the experimental requirements.

Su et al~\cite{Su} proposed the uniform weight approach. The problem
that they tackle is of choosing sites for placement of antennas
given several more candidate sites. They aim for a uniform and
complete UV-distribution in the absence of a-priori information on
the objectives of the experiments. Their method involves assigning a
weight to a UV point that is equal to its distance from the nearest
UV point. An antenna is thus assigned the sum of all distances of
the relevant UV points. The antenna that has the least weight is
dropped.

Finally, we mention the tomographic projection method of de
Villiers~\cite{deVilliers} in which he compares the projection of an
imperfect and an ideal UV distributions onto one dimension. The
discrepancy yields correction terms that are mapped to new antenna
positions. Doing this in several directions allows one to get close
to the ideal configuration.

\section{The Minimum Variance Method} \label{sec:mvm}
The basic idea of this paper is as follows. First, we divide the
UV-plane into $p$ regions. Division is necessary because we wish a
distributed removal of points so that some portion of the UV plane
is not denuded of UV points. The general way to capture the notion
of distribution is to grid the plane and then talk of ``points per
grid unit"~\cite{Boone2001,Su}. How this division is done will be
described towards the end of this section. We note, however, that
the MVM can accommodate any scheme for division of the UV plane in
principle. The desired UV distribution dictates the number of UV
points each region should hold.

Our algorithm starts with $M > N$ antennas and removes $M - N$
antennas, one at a time, in an iterative fashion. Suppose that at
the start of the $r^{\tiny\hbox{th}}$ iteration, there are $p^{[r]}$
regions and $M^{[r]}$ antennas. Then, $M^{[r]}-1$ UV points will be
dropped as a result of the removal of one antenna during the
$r^{\tiny\hbox{th}}$ iteration.

Depending on the requirements of the UV distribution, we would like
the removal of $R_{i}^{[r]}$ UV points from the $i^{\tiny\hbox{th}}$
region in the $r^{\tiny\hbox{th}}$ iteration. However, in general,
the number of UV points dropped from the region $i$ due to removal
of antenna $j$ would be different, say $w_{i}(j)$.

We capture the discrepancy between the actual and ideal UV points
dropped in every region by removing the antenna $j$ with the
following term:
\[
Var(j) = \sum_{i=1}^{p^{[r]}} (w_{i}(j)-R_{i}^{[r]})^2.
\]
We now drop the antenna that has the least $Var(.)$ value. In short,
the logic for deletion of the $N$ antennas is this:

\begin{enumerate}

\item \quad\quad \emph{while there are more than $N$ antennas remaining do }
\item \quad \quad\quad \emph{for each remaining antenna $j$}
\item \quad \quad\quad \quad Calculate $Var(j)$.
\item \quad \quad\quad \emph{end for}
\item \quad \quad\quad Remove the antenna that has the least $Var(.)$.
\item \quad \quad\emph{end while}
\end{enumerate}

This antenna removal routine can be tweaked for different problems.
Let us now estimate the worst case time complexity of this routine.
There are at most $M-N$ antennas to be removed. Every remaining
antenna is a candidate for removal, of which there are at most $M$.
For every such candidate we go through the (at most $M^2$) UV
points, identify those associated with the candidate, and determine
which region they belong to, in order to calculate $Var(j)$. Thus,
we need at most $O((M-N)M(M^2+p))=O((M-N)\cdot M^3)$ steps.

But for a rotation for changing coordinates, a UV point is defined
as $u = \frac{X_1-X_2}{\lambda}$ and $v = \frac{Y_1-Y_2}{\lambda}$
for every pair $(X_1,Y_1)$ and $(X_2,Y_2)$ of XY coordinates (see
Appendix A). Therefore it is sufficient to work with the
distribution that the XY coordinate differences may follow. In all
our experiments, without loss of generality, we take $\lambda=1$.

There are at least two natural choices for the shape of the regions
into which the UV plane can be divided. One is into concentric
circles due to radial symmetry. The other is into coaxial ellipses
because of the facts that the UV tracks are elliptical in general,
and that in several earth rotation aperture synthesis telescopes, a
prolonged coverage is desirable. In a scheme that removes antennas
progressively from a large set, there are two options regarding  the
number of regions. The first is to adaptively reduce the number of
regions in accordance with the reduced antennas. In doing so, the
configuration in an iteration will mimic that in the previous
iteration. The second option is to stick to the same number of
regions throughout, so that all intermediate configurations will
mimic the final. We demonstrate the use of different UV plane
division schemes, both in terms and shape and number of regions, in
our experiments.

\section{Applications}  \label{sec:app}
\subsection{Problem 1: UV to XY}    \label{subsec:prob1}

The UV distribution that is desired in most cases is a Gaussian
along the radial and uniform along the azimuthal direction. As the
next lemma indicates, a random distribution of antennas on the
ground, that is, the XY-plane, gives a tapering distribution in the
UV-plane.

\begin{bf}Lemma: \end{bf}
Consider a $S\times S$ square grid having $(S+1)^2$ lattice points
in the XY plane. Suppose that we choose each lattice point
independently with probability $0 \leq p \leq 1$ and look at the UV
distribution generated by the chosen points. Then the expected
number of UV points at the position $(x,y)$, where $-S \leq x, y
\leq S$, is $p^2(S+1-|x|)(S+1-|y|)$.

\begin{bf}Proof: \end{bf}
For the given $S\times S$ square grid, the number of pairs of
lattice points $(X_1,Y_1)$ and $(X_2, Y_2)$ such that $X_2 - X_1 =
x$ and $Y_2 - Y_1 = y$ is $(S+1-|x|)(S+1-|y|)$. Since each such
contributing pair is expected to be chosen with probability $p^2$,
linearity of expectation gives the desired result. \hfill $\square$

The above lemma indicates that a random XY distribution induces more
UV points corresponding to short XY distance pairs. This tapers off
for longer distances. With this as a starting point, we apply the
Minimum Variance Method to arrive at the target Gaussian.

\medskip

\noindent{\bf Given}: (i) A geography, that is, dimensions of the
land on which to place antennas. (ii) The number of antennas that
are to be installed and (iii) The amplitude $a_f$ of the desired
Gaussian UV plane.

\noindent {\bf Goal}: To find a placement of $N$ antennas that
yields a Gaussian UV distribution centered at the origin having an
amplitude $a_f$ and standard deviation $\sigma=B/4$, where $B$ is
the largest inter UV point distance.\footnote{We work with
$\sigma=B/4$, since the area under the gaussian distribution curve
before the $4\sigma$ coordinate closely approximates the total
area.}

\medskip

\noindent{\bf Approach}: We start with an overkill of antennas.
Instead of the required $N$, we start with $M \approx 2N$ antennas
and place them randomly on the XY-plane. Our random choice results
in a distribution of points on the UV-plane that deviates from the
desired Gaussian. We now try to minimize the deviation from the
target Gaussian by iteratively removing an antenna that causes the
minimum deviation from the desired Gaussian. We stop when we have
removed $M - N$ antennas.

\begin{figure}\label{curves}
\centering
\includegraphics[width=0.5\textwidth]{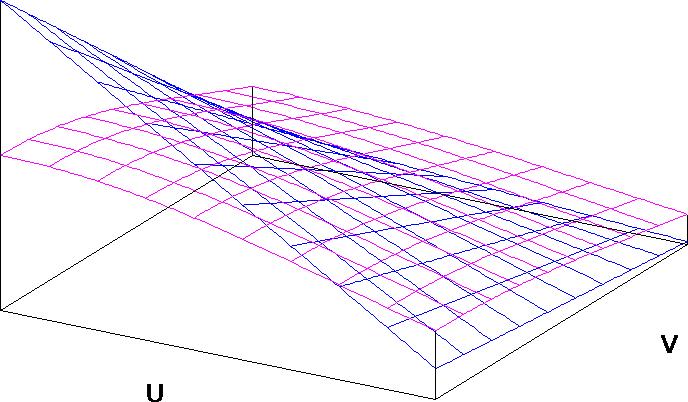}
\caption{The steeper curve is generated by an initial random
distribution of the $M$ antennas while the pink curve is the
desired Gaussian.} %
\label{fig:gaussian}
\end{figure}

Figure \ref{fig:gaussian} illustrates conceptually the line of
approach. The two meshes indicate the starting and the final
distributions. The effort will be towards minimizing the discrepancy
between the distributions at every iteration when we remove an
antenna.

\medskip
\noindent{\bf Details}: As stated earlier, our random choice results
in a distribution of points on the UV-plane that deviates from the
desired Gaussian.

We then draw a fixed number $p=B/d$ of concentric circles with the
radii increasing in arithmetic progression of common difference $d$.
We will show how to fix $d$ shortly. Given these annuli, it is
possible to plot a \emph{density} histogram of the number
$q_i^{[\tiny\hbox{init}]}$ of UV-points that lie in the annuli
bounded by the circles $i$ and $i+1$ (for $1<i<p$;
$q_1^{[\tiny\hbox{init}]}$ being the number of points in the central
circle) versus distance. Thus, the total number of UV points
initially is $Q^{[\tiny\hbox{init}]}=\sum_{i=1}^{p}
q_i^{[\tiny\hbox{init}]}=M\cdot (M-1)/2$ and area under this
histogram is $A^{[\tiny\hbox{init}]}=d\cdot\sum_{i=1}^{p}
q_i^{[\tiny\hbox{init}]} = d \cdot M\cdot(M-1)/2$.

Let us now fix $d$. The area under half the Gaussian centered at the
origin with amplitude $a$ and having a standard deviation of
$\sigma$ is given by $a\sigma \sqrt{\frac{\pi}{2}}$ (see Appendix
B). Thus, given $a_f$, $\sigma$ and the final number $N$ of
antennas, we have $a_f\sigma\sqrt{\frac{\pi}{2}}=\frac{d\cdot
N\cdot(N-1)}{2}$. Therefore,
$d=\frac{a_f\sigma\sqrt{\frac{\pi}{2}}}{\frac{N\cdot(N-1)}{2}}=\frac{a_fB\sqrt{\frac{\pi}{2}}}{2N\cdot(N-1)}$.

As has been mentioned previously, the central idea behind the
present approach is to crop the initial curve to fit the Gaussian.
Let the number of antennas remaining at the end of the $r^{th}$
iteration ($1 \leq r \leq M-N$) be $M^{[r]} = M - r$. Then, $M^{[0]}
= M$. Note that removal of an antenna during the $r^{th}$ iteration
results in the removal of $M^{[r-1]}-1$      
UV-points, which in turn
results in a reduction of the area under the curve from $A^{[r-1]}$
to $A^{[r]} = d \cdot M^{[r]} \cdot (M^{[r]}-1)/2$.

We try to minimize the deviation from Gaussian by iteratively
removing the antenna that causes the maximum deviation from the
desired Gaussian.  We use $ A^{[r]}  =  d\cdot M^{[r]} \cdot
(M^{[r]}-1)/2 = a\sigma \sqrt{\frac{\pi}{2}}$, to get $a  =
\frac{d\cdot(M-r)\cdot(M-r-1)/2}{\sigma \sqrt{\frac{\pi}{2}}}. $

With $a$ determined (and $\sigma$ already being fixed), the expected
Gaussian is uniquely defined. The vertical coordinate in the
Gaussian distribution at $x=i$ is given by the formula
$ae^{-\frac{d\cdot(i-1/2)^2}{2\sigma^2}}$. If $q_i^{[r-1]}$ is the
number of UV points remaining in region $i$ at the end of iteration
$r-1$, then the number of UV points we would like to remove from
$i^{\tiny\hbox{th}}$ region in $r^{\tiny\hbox{th}}$ iteration is
\begin{eqnarray*}
R_{i}^{[r]} & = & q_i^{[r-1]} -
\frac{d\cdot(M-r)\cdot(M-r-1)/2}{\sigma\sqrt{\frac{\pi}{2}}} \cdot
e^{-\frac{d\cdot(i-1/2)^2}{2\sigma^2}}.
\end{eqnarray*}

For each region $i$, let $w_{i}(j)$ be the number of UV-points that
would be deleted from the region $i$ if the antenna $j$ is removed.
Then, as per the Minimum Variance Method, we choose that antenna $j$
for removal which has got the least value of
\[
Var(j) = \sum_{i=1}^p  (w_{i}(j) - (q_i^{[r-1]} -
\frac{d\cdot(M-r)\cdot(M-r-1)/2}{\sigma\sqrt{\frac{\pi}{2}}} \cdot
e^{-\frac{d\cdot(i-1/2)^2}{2\sigma^2}}))^2.
\]

\begin{figure}[ht]
\centering %
\subfigure[The initial set of antennas]{
\includegraphics[width=2.5in]{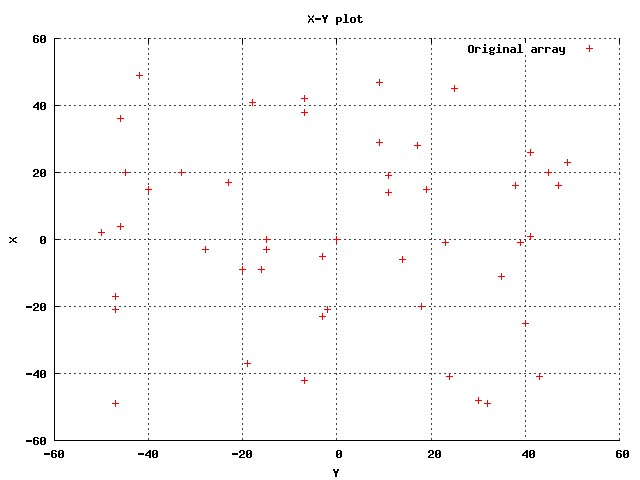}
\label{fig:subfig1} %
}%
\subfigure[The final set of antennas]{
\includegraphics[width=2.5in]{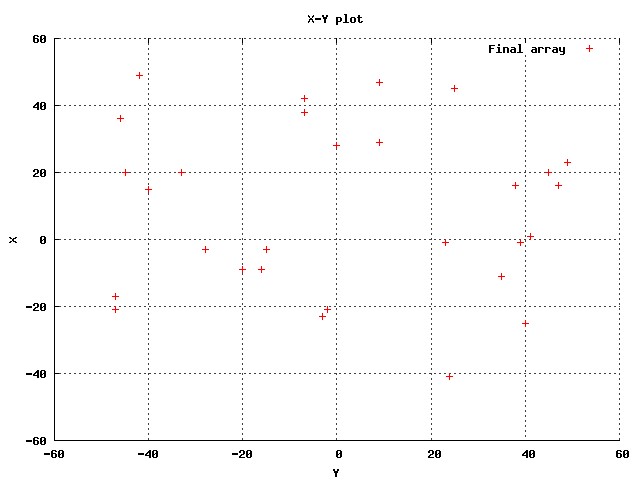}
\label{fig:subfig6} %
}%
\end{figure}

For the experiment, we set a goal of generating a Gaussian
distribution on the UV plane using $N=28$ antennas with $a_f=70$. We
start with an initial set of $M=48$ antennas that are generated
randomly as shown in Figure \ref{fig:subfig1}. The geography that we
use, and the randomly generated antenna positions yields
$B\approx11180$ wavelengths. Then, $d \approx6470$ wavelengths.

\begin{figure}[ht]
\centering %
\subfigure[The initial UV-distribution]{
\includegraphics[width=2.5in]{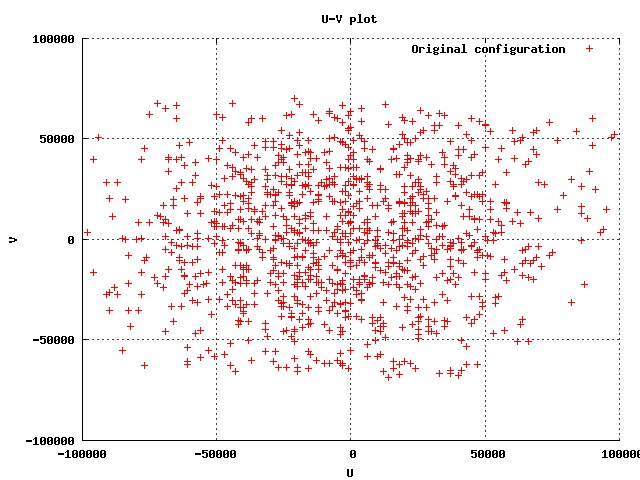}
\label{fig:subfig2}
}%
\subfigure[The final UV-distribution]{
\includegraphics[width=2.5in]{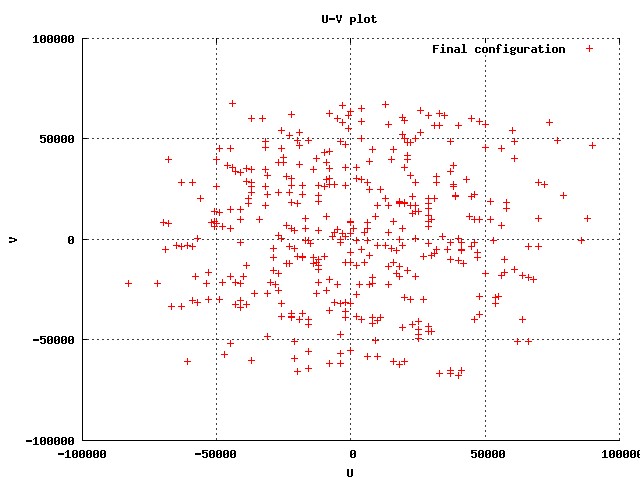}
\label{fig:subfig5}
}%
\end{figure}

\begin{figure}[ht]
\centering %
\subfigure[The initial UV-density histogram]{
\includegraphics[width=2.5in]{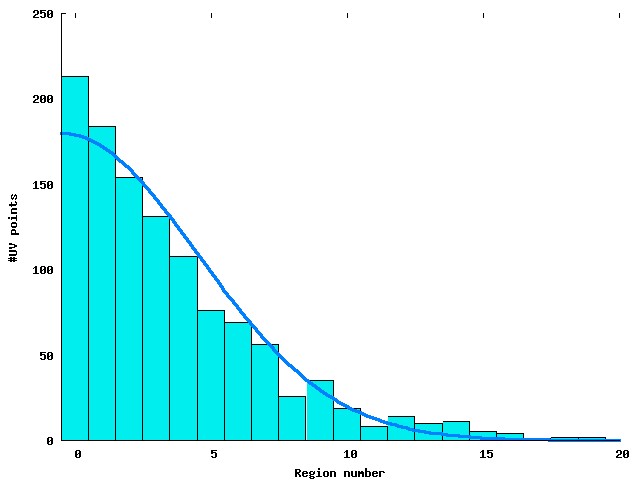}
\label{fig:subfig3}
}%
\subfigure[The final UV-density histogram]{
\includegraphics[width=2.5in]{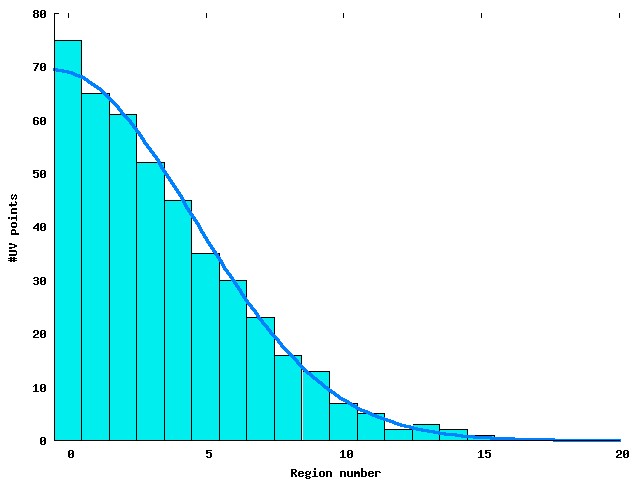}
\label{fig:subfig4}
}%
\label{fig:rand_exp} %
\caption{UV to XY}
\end{figure}

Figure \ref{fig:subfig2} shows the UV-distribution generated by
these antennas. The Density Histogram of the distribution is shown
in Figure \ref{fig:subfig3}. Notice that the histogram deviates from
the desired Gaussian at several places.

Figure \ref{fig:subfig4} is the cropped histogram arrived at
finally. Observe that this histogram is closer to Gaussian than the
initial one. Figures \ref{fig:subfig5} and \ref{fig:subfig6} show
the final UV and XY distributions respectively.

\subsection{Problem 2: Staggered Construction} \label{subsec:prob2}
\emph{Statement}: A telescope of $M$ antennas has been proposed with
all the sites identified. However, the telescope construction has to
be staggered and it has to be built in phases since the funding
becomes available in small chunks over a time period. Suppose that
we are given a budget of $N$ antennas for Phase 1. Which $N$ of the
total $M$ antennas should we construct in Phase 1?

Ideally one would like to start making quality observations after
completion of Phase 1 itself which means that the UV-distribution
that is generated by Phase 1 antennas should be a good approximation
of the final UV-distribution that we would get after placing all $N$
antennas.

If $M$ is even, then we choose $p^{[0]}= M-1$ ellipses with $M/2$
UV-points per ellipse. Else, we choose $p^{[0]} = M$ ellipses with
$(M-1)/2$ UV-points per ellipse. One ellipse is chosen for every $M$
of all the elliptical UV-tracks sorted by their
$u^2+\frac{v^2}{\sin^2\delta}$. We aim for the removal of the same
number of UV points from each region during every iteration. Thus,
$R_{i}^{[r]}=\frac{M^{[r]}-1}{p^{[r]}}$, and
\[
Var(j) = \sum_{i=1}^{p^{[r]}}
(w_{i}(j)-\frac{M^{[r]}-1}{p^{[r]}})^2.
\]

\subsubsection{Experimental Results} \label{subsec:express}
\begin{figure}[!h]
\centering \subfigure[XY plane--initial and 40 removed antennas]{
\includegraphics[width=2.2in]{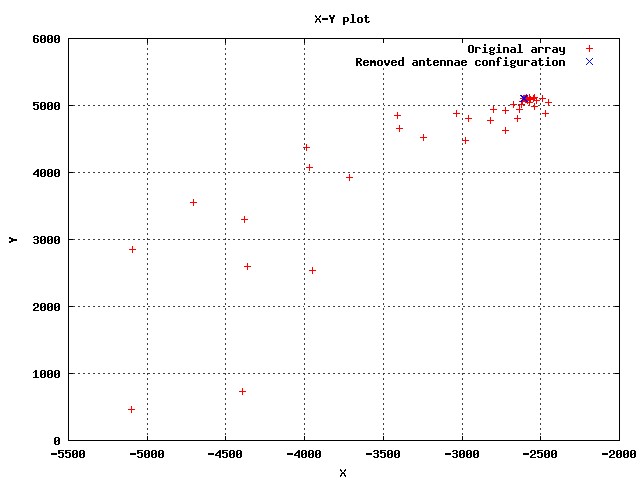}
\label{fig:xyplot-aus40} } \subfigure[XY plane--initial and 80
removed antennas ]{
\includegraphics[width=2.2in]{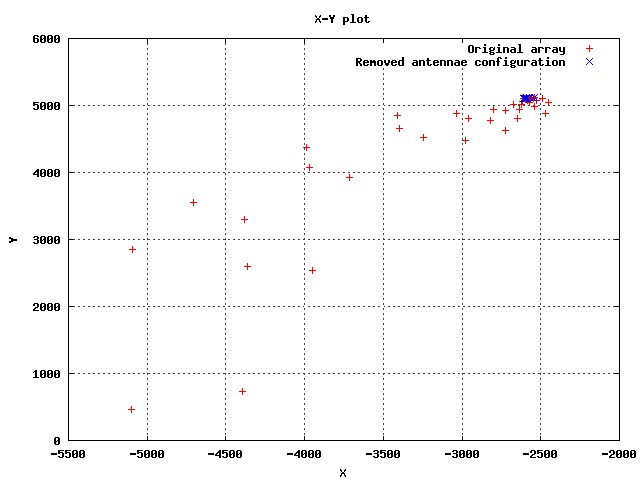}
\label{fig:xyplot-aus80} } \subfigure[XZ plane--initial and 40
removed antennas]{
\includegraphics[width=2.2in]{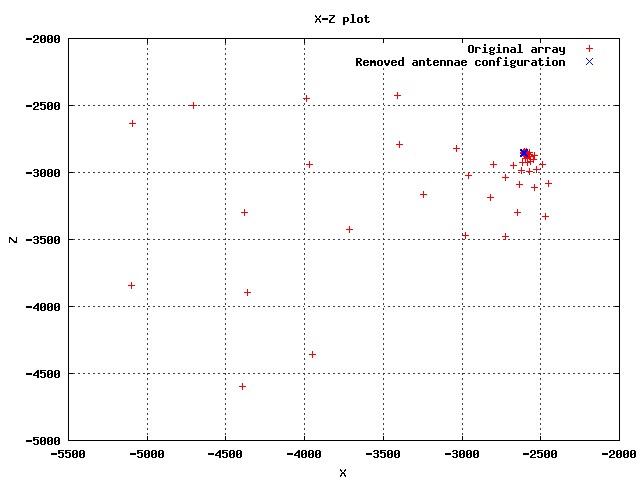}
\label{fig:xzplot-aus40} } \subfigure[XZ plane--initial and 80
removed antennas ]{
\includegraphics[width=2.2in]{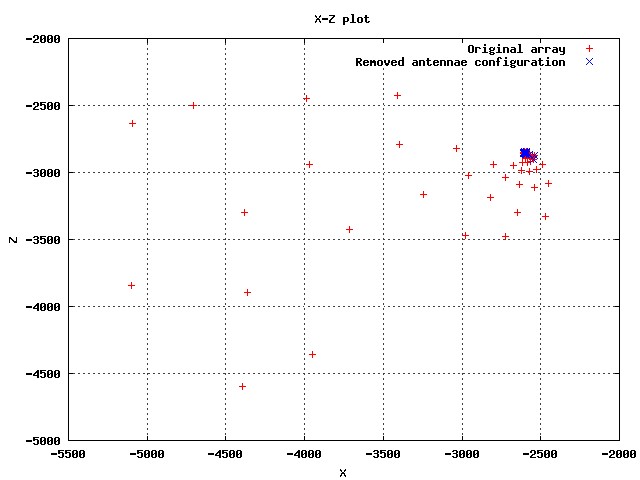}
\label{fig:xzplot-aus80} } \subfigure[The UV plane--initial and 40
removed antennas]{
\includegraphics[width=2.2in]{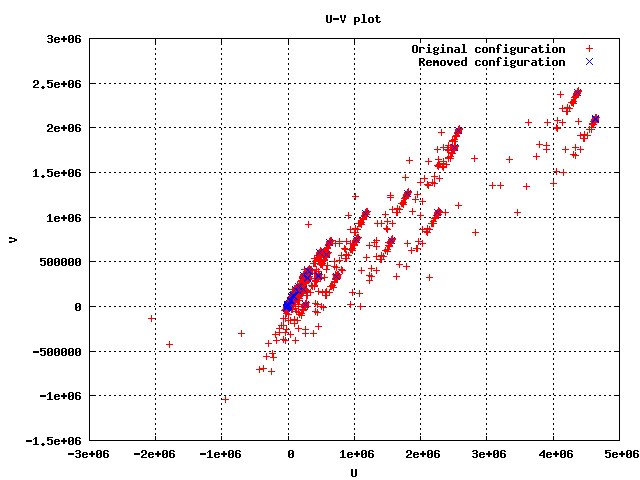}
\label{fig:uvplot-aus40} } \subfigure[The UV plane--initial and 80
removed antennas]{
\includegraphics[width=2.2in]{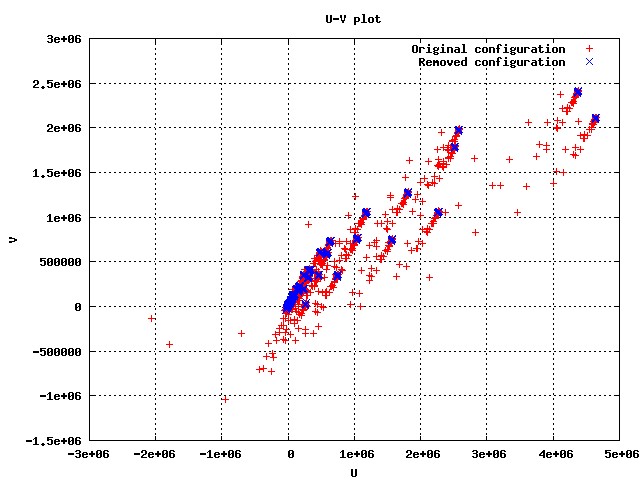}
\label{fig:uvplot-aus80} } \caption{Results for the proposed
Australian SKA configuration.}
\end{figure}

We ran the experiment on the scaled down version of the Australian
SKA for $M=120$ antenna locations. Figures \ref{fig:xyplot-aus40}
and \ref{fig:xyplot-aus80} show the XY plane when populated with all
the $120$ antennas and after removal of $40$ antennas and $80$
antennas respectively.

Since the $Z$ coordinates of the antennas also contribute to
defining the UV plane, we provide the corresponding XZ plots along
with the XY plots. Figures~\ref{fig:xzplot-aus40}
and~\ref{fig:xzplot-aus80} show the corresponding XZ planes.

Figures \ref{fig:uvplot-aus40} and \ref{fig:uvplot-aus80} show the
corresponding UV-planes. The hour angle used is $0^\circ$ and the
declination $-30^\circ$.

Note that by our scheme, more antennas are removed towards the
center of the log-spiral. Thus, it is more effective when $N$ is
large. Also observe that by making use of the same Minimum Variance
Method, we can create a construction plan for all remainder phases
of the telescope building which has a steadily improving
UV-coverage.
\subsection{Problem 3: Mobile Antennas} \label{subsec:prob3}
\emph{Statement}: Suppose there are $L$ pads (for mounting $L$
antennas), but only $M\leq L$ mobile antennas pre-placed on $M$
pads. Suppose that an astronomer requests these $M$ antennas for
some observation. However, only $N$ out of these $M$ can be
allocated (say, because of maintenance reasons, or some being needed
for another experiment). Our problem is to choose $N$ out of the $M$
antenna pads that best approximate the desired UV pattern generated
by fully functional pre-placed $M$ antennas. Having done that, the
antennas can be shifted to the recommended $N$ pads.

\begin{figure}[!h]
\centering %
\subfigure[The XY plane]{
\includegraphics[width=2.2in]{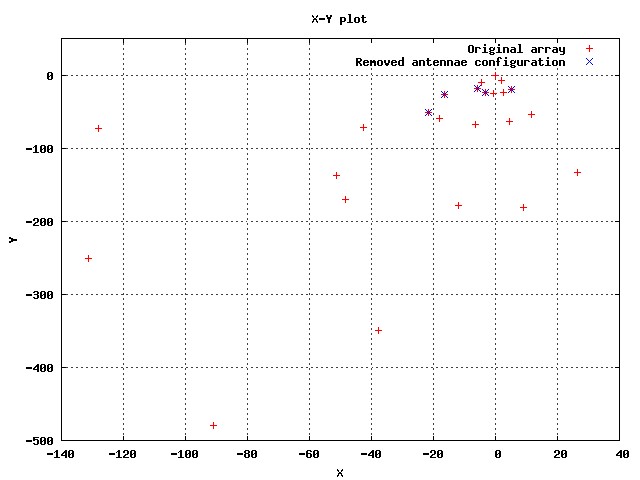}
\label{fig:harv_xy} %
}
\subfigure[The XZ plane]{
\includegraphics[width=2.2in]{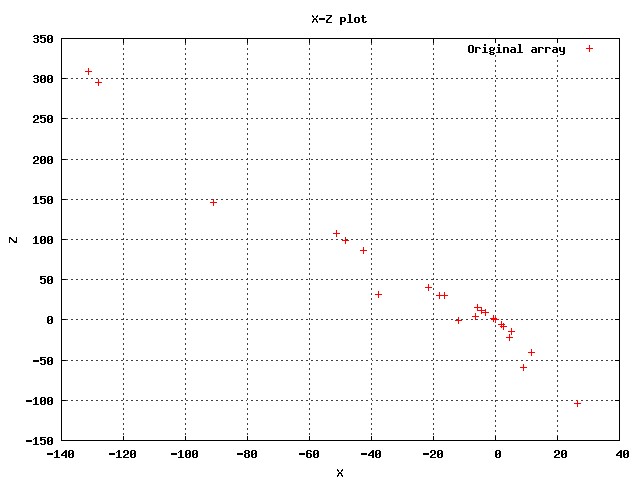}
\label{fig:harv_xz} %
}
\subfigure[The UV plane]{
\includegraphics[width=2.2in]{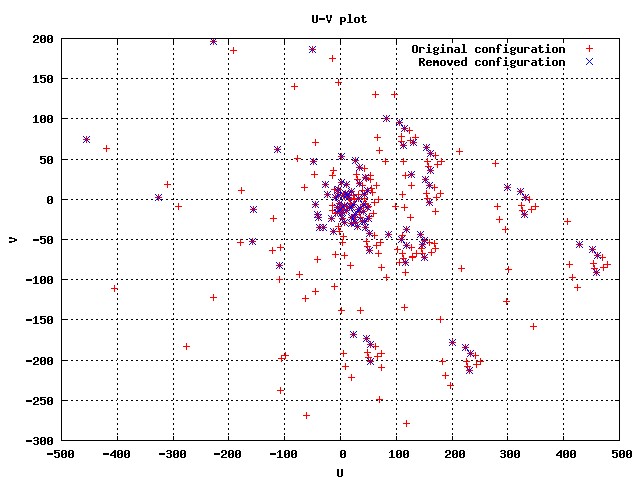}
\label{fig:harv_uv} %
}%
\caption{Results for SMA configuration (mobile antennas)}
\end{figure}

The parameters set for the previous problem in Subsection
\ref{subsec:prob2} carries through exactly. We did the experiment on
Sub Millimeter Array data~\cite{Karast}. We assume the failure of
$5$ of the $24$ antennas. Thus, in this setting, $M=24$ and $N=19$.
Figure \ref{fig:harv_xy} shows the initial and final XY planes and
\ref{fig:harv_xz} shows the XZ plane. Figure \ref{fig:harv_uv} show
the resultant UV distributions. Using the output of the algorithm,
one can recommend which $19$ pads can be used.
\section{Conclusions and Future Work}   \label{sec:concl}
In this paper we proposed and studied a Minimum Variance Method that
can be used in many radio antenna placement scenarios. We applied
the method in three different situations effectively (i) obtaining
ab-initio smooth Gaussian UV distributions (ii) incremental
construction of very large aperture synthesis arrays and (iii)
achieving fault tolerance in mobile antennas. We also report
experiments that indicate the usefulness of this method. In addition
versatility, this method is quite efficient when compared to the
brute force method or minor improvements thereof.

In this paper, the only criterion that we considered for placing the
antennas is image quality. It remains to come up with solutions that
take into account logistical factors like wire length minimization,
roadway utilization and so on.

\newpage
\appendix
\section{Aperture Synthesis Primer} \label{app:primer}
It is a well-known fact in astronomy that the angular resolution of
a telescope is proportional to $\lambda/D$ where $\lambda$ is the
wavelength of the waves to be observed, and $D$ is the diameter of
the telescope aperture. In order to observe celestial bodies in the
long wavelength ranges ($1 \text{mm}$ to $10\text{m}$), we need
antennas with diameters of the order of hundreds of kilometers.
Obviously, such antennas are practically impossible to build,
manoeuver and maintain. Fortunately, it has been shown that several
antennas can be used instead of a single big one. Several radio
telescopes have been constructed in the past based on this principle
of \emph{aperture synthesis}. These telescopes have helped
astronomers in discovering various celestial bodies like quasars,
pulsars etc.

In this Appendix we quickly recap some concepts and terminology
pertinent to this paper. The interested reader is referred
to~\cite{Burke} and~\cite{CGD} for an excellent treatment of the
subject.

Let us first consider the case of two antennas. Consider a
rectilinear coordinate system $(u,v,w)$ such that the $w$-axis
points in the direction that is to be the center of the synthesized
field of view. Let $\mathbf{s_0}$ be the unit vector along the
w-axis.

Assigning a direction and treating the shortest distance between the
two antennas as the magnitude, we can speak of a baseline
vector.\footnote{In general, a long baseline improves the resolution
of the array, while a short baseline implies a larger field of view.
A larger number of different baselines implies higher sensitivity.}
Let the baseline vector be $\mathbf{b}=u\hat{u} +v\hat{v} +
w\hat{w}$. Let $l$, $m$ and $n$ be the direction cosines of an
element in the celestial sphere. Let $\mathbf{s}=l\hat{u} + m\hat{v}
+ n\hat{w}$ be a position vector of another element close by. Then,
$\mathbf{b.s}= ul+vm+wn$ and $\mathbf{b.s_0}=w$.

Let $A(l,m)$ be the response of the antennas corresponding to a
baseline in the direction specified by $(l,m)$. Further, let $A_0$
be the response of the antennas at the beam center. Then,
$\mathcal{A}(l,m)=\frac{A(l,m)}{A_0}$ is called the normalized
antenna reception pattern. The spatial correlation function of the
electric field at the antennas corresponding to the baseline
$\mathbf{b}$ is defined as
\[
V(u,v,w)=
\int_{-\infty}^{\infty}\int_{-\infty}^{\infty}\mathcal{A}(l,m)I(l,m)e^{-2\pi
i[ul+vm+w(\sqrt{1-l^2-m^2}-1)]} \frac{dldm}{\sqrt{1-l^2-m^2}}
\]
where $I(l,m)$ is the radio brightness in the $(l,m)$ direction,
assuming that both the antennas are identically polarized.

Obtaining $I(l,m)$, our aim, is simplified to evaluating the inverse
of a 2-D Fourier transform if $w=0$ :
\[
V(u,v)
=\int_{-\infty}^{\infty}\int_{-\infty}^{\infty}A(l,m)I(l,m)e^{-2\pi
i[ul+vm]} \frac{dldm}{\sqrt{1-l^2-m^2}}
\]

Therefore, we sample a plane perpendicular to $\hat{w}$, the
direction of our interest. Every pair of antennas generates a point
on the UV-plane will be called a uv-point.

\begin{figure}
\centering
\includegraphics[width=2.5in]{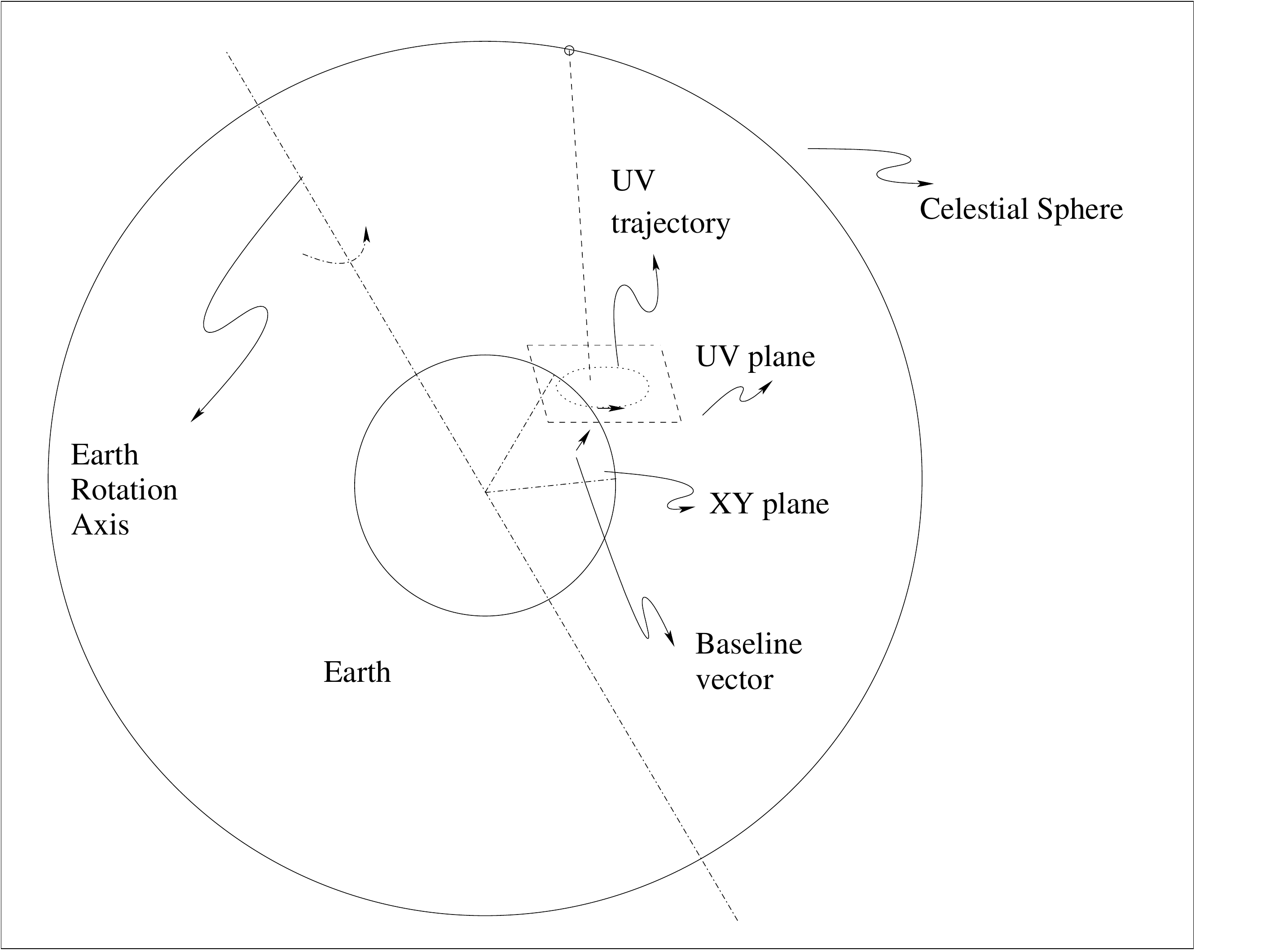}
\caption{The XY and UV planes} \label{fig:expl}
\end{figure}

In the case of a large number of antennas, a convenient coordinate
system is used for the XY-plane~\cite{NRAO}.  We choose the $Z-$
direction to be along the north pole, the $X-$ axis and the $Y-$
axis lie on the equatorial plane with the $X-$ axis pointing in the
direction of the Greenwich meridian and the $Y-$ axis pointing to
the East, and the origin (0,0,0) lies at the center of the earth.
Figure~\ref{fig:expl} shows the XY and UV planes in some detail.

The following relates the latitudes and longitudes of a point on the
earth to its X and Y coordinates in the system defined above.
\begin{eqnarray}
X&=&R_e\cos(\theta_{lat})\cos(\theta_{long})\nonumber\\
Y&=&R_e\cos(\theta_{lat})\sin(\theta_{long})\nonumber\\
Z&=&R_e\sin(\theta_{lat})\nonumber,\\ \nonumber
\end{eqnarray}
where $R_e$ is the radius of the earth, and $\theta_{lat}$ and
$\theta_{long}$ are the latitude and longitude of the location
respectively.

Consider now two antennas placed at $(X_1,Y_1,Z_1)$ and
$(X_2,Y_2,Z_2)$. Let $L_Y= X_2-X_1$, $L_Z=Y_2-Y_1 $ and
$L_X=Z_2-Z_1$.

Then, the $(u,v,w)$ coordinates are related to $L_X,L_Y,L_Z$ as
follows:

\begin{equation}
\left(\begin{array}{c} u\\v\\w\\\end{array}\right)=
\frac{1}{\lambda}\left(\begin{array}{ccc} \sin H & \cos H & 0\\
-\sin \delta cos H & \sin \delta sin H & \cos \delta \\
\cos \delta cos H & -\cos \delta sin H & \sin \delta \\
\end{array}\right)
\left(\begin{array}{c} L_X\\L_Y\\L_Z\\\end{array}\right)
\end{equation}
where $H$ is the hour angle and $\delta$ is the declination of the
observation direction respectively, and $\lambda$ is the wavelength.
While the curve traced out by a single baseline is coplanar,
different baselines will lie on parallel planes, but at different
elevations. The equation of a UV-curve is given by
\[
u^2 + \left(\frac{v-(L_Z/\lambda)\cos\delta}{\sin\delta}\right)^2 =
\frac{L^2_X+L^2_Y}{\lambda^2}.
\]

\begin{figure}[ht]
\centering %
\subfigure[UV track for $L_Z = 0$]{
\includegraphics[width=2.5in]{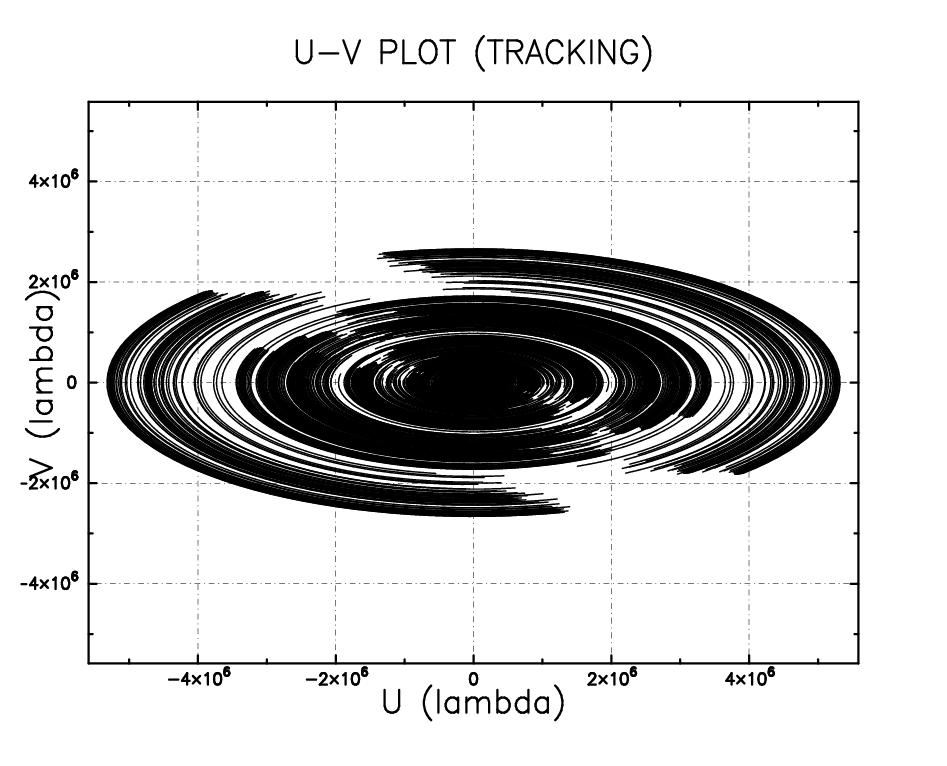}
\label{subfig:uvtrck0} %
}%
\subfigure[UV track for $L_Z \ne 0$]{
\includegraphics[width=2.5in]{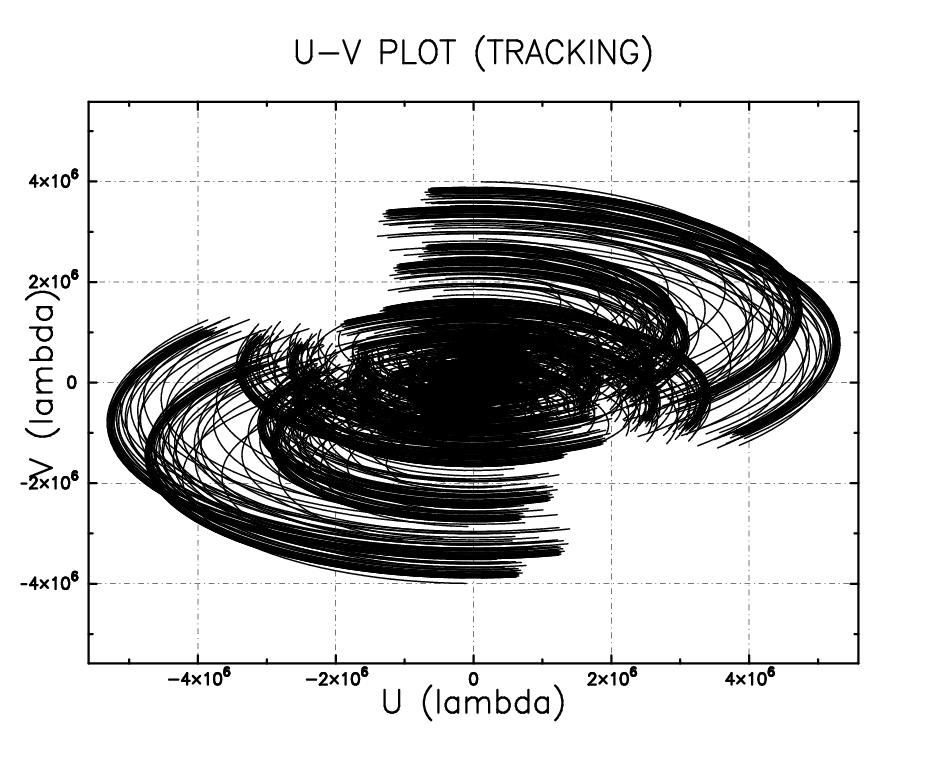}
\label{subfig:uvtrck} %
}%
\caption{UV tracking for Australian SKA proposed antennas/stations
locations for declination = $-30^o$}
\end{figure}

The equation clearly shows that for given values of $L_X$, $L_Y$ and
$L_Z$, the path trace by a UV point is an ellipse. If the $L_Z=0$,
the origin will be the center of the ellipse. For $L_Z \ne 0$ the
ellipse center is shifted accordingly on $v$ axis by $(L_Z/\lambda)
\cos\delta$. For example, see Figure \ref{subfig:uvtrck0} and
\ref{subfig:uvtrck}, plotted for Australian proposed SKA
antennas/stations.

We finish by stating some useful definitions and remarks. The
earth's axis determines half planes with the zenith and the source
respectively. The angle between these two planes is called the
\emph{hour angle}. The angular distance of a source above or below
the celestial equator is called its \emph{declination}.

For good image quality, it is desirable that we sample as many
points on the UV-plane as possible. As the earth rotates, the
UV-points trace out curves on the UV-plane, populating it. This
technique of making use of the earth's rotation to fill up the
sampling plane is called \emph{aperture synthesis}. Naturally, we
require that over time, the UV-plane is filled up as much as
possible.

It is a useful exercise to work out the curve traced by a UV-point
as the earth rotates. In particular, one will observe that the curve
traced for a source at 0 declination is a straight line, that for
$90^\circ$ declination is a circle, while all intermediate sources
trace out ellipses of increasing eccentricity.

While different sources would require different UV distributions, it
has been suggested  that (i) in case of a dense UV plane, a Gaussian
distribution is preferred in the radial direction, and a uniform
distribution is preferred across the azimuth; (ii) in case of a
sparse UV plane, a uniform distribution over the plane is
preferred~\cite{Boone2002}  .

\section{Area Under the Gaussian}

Consider a Gaussian $y=a e^{-\frac{x^2}{2\sigma^2}}$ centered at
$x=0$, amplitude $a$ and variance $\sigma^2$. Denote by $A$ the area
under the curve: $A=a \int_{-\infty}^{\infty}
e^{-\frac{x^2}{2\sigma^2}}dx$. Then, $A^2 =a^2
\big\{\int_{-\infty}^{\infty} e^{-\frac{x^2}{2\sigma^2}}dx\big\}^2$.
Or, $A^2 =a^2 \int_{-\infty}^{\infty}\int_{-\infty}^{\infty}
e^{-\frac{x^2+y^2}{2\sigma^2}}dxdy$.

Writing it in polar coordinates, we get $A^2=
a^2\int_0^{2\pi}d\theta\int_0^{\infty}
e^{-\frac{r^2}{2\sigma^2}}rdr$. This evaluates to $2\pi a^2c^2$.
Therefore, $A=\sqrt{2\pi}ac$.
%
%
\end{document}